\newcommand{\stitolo}{Leptons with E$>$200 MeV trapped in the Earth's radiation 
belts }  

\newcommand{\sautori}{

E. Fiandrini$^{\mbox{a}}$,            
G. Esposito$^{\mbox{a}}$,         
B. Bertucci$^{\mbox{a}}$,             
B. Alpat$^{\mbox{a}}$,             
R. Battiston$^{\mbox{a}}$,             
W.J. Burger$^{\mbox{a}}$,             
G. Lamanna$^{\mbox{a}}$,             
P. Zuccon$^{\mbox{a}}$             

}

\newcommand{\sistituzioni}{
                                      
$^{\mbox{a}}$                       
University and INFN of Perugia - Italy  }  

\newcommand{\ssubmitted}{

Submitted to Journal of Geophysical Research    

}
\newcommand{\sabst}{

For the first time accurate measurements of electron and positron fluxes in the energy range 0.2$\div$10 GeV
have been performed with the Alpha Magnetic Spectrometer (AMS) at altitudes of 
370$\div$390 km in the geographic latitude interval $\pm 51.7^o$.
We describe the observed under-cutoff lepton fluxes outside the region
of the South Atlantic Anomaly (SAA). The separation in {\it quasi-trapped}, long lifetime (O(10 s)), 
and {\it albedo}, short
lifetime (O(100 ms)), components is explained in terms of the drift shell
populations observed by AMS. A significantly higher relative abundance of positrons
with respect to electrons is seen in the {\it quasi-trapped} population.
The flux maps as a function of the canonical adiabatic variables L, $\alpha_{o}$ are presented 
for the interval 
$0.95\!<L<\!3$, $0^{\mathrm{o}}\!\!<\alpha_{o}\!<\!90^{\mathrm{o}}$  
for electrons (E$<$10 GeV) and positrons (E$<$3 GeV). The results are compared with existing data at 
lower energies.
The properties of the observed under-cutoff particles are also investigated in terms of their
residence times and geographical origin.

}
\documentclass[12pt,twoside]{article}
\usepackage{epsfig}                     
\newcommand{\titolo}{
\begin{center}\Large{\bf{
\stitolo
}}\end{center}}
\newcommand{\autori}{
\begin{center}
\sautori
\end{center}}
\newcommand{\istituzioni}{
\begin{center}
\small{\it{
\begin{tabular}{c}                    
\sistituzioni
\end{tabular}}}                       
\end{center}}                         
\newcommand{\submitted}{
\begin{center}
\small{\it{
\ssubmitted
}}
\end{center}}
\newcommand{\abst}{
\begin{abstract}
\sabst               
\end{abstract}}
\begin{document}
\vfill
\titolo
\vfill
\autori
\istituzioni
\vfill
\submitted
\abst
\vfill
\clearpage     
\setcounter{page}{2} 
\pagestyle{plain}                
%


%
\section{Introduction}                

Evidence for high energy (up to few hundred MeV) electrons and positrons trapped below 
the Inner Van Allen Belts has been published during the last 20 years.
The existing experimental data in the energy range of 0.04$\div$200 MeV come from satellites covering a 
large range of adiabatic variables [\cite{Voronov87}, \cite{Galper86}, 
\cite{Akimov87},\cite{Heynderickx97}].
Additional information, at relatively higher energies, is furnished by balloon-borne experiments 
[\cite{Verma67},\cite{Barwick98}]; 
however these data cover a more limited spatial range and have larger uncertanties due to the
shorter exposure times and the presence of background from atmospheric showers.
Although the magnetic trapping mechanism is well understood, a complete description of the phenomena,
including the mechanisms responsable for the injection and depletion of the belts as well as those determining the energy spectra is lacking,
particularly for energies above a few hundred MeV.
At lower energies,  models are  available for leptons and protons
[\cite{Vette91},\cite{Getselev91}] based
on the data provided by satellite cam\-pai\-gns, which are continuosly updated for istance in the context
of the Trapped Radiation ENvironment Development project [\cite{TREND}].

\begin{figure}
\begin{center}
\epsfig{file=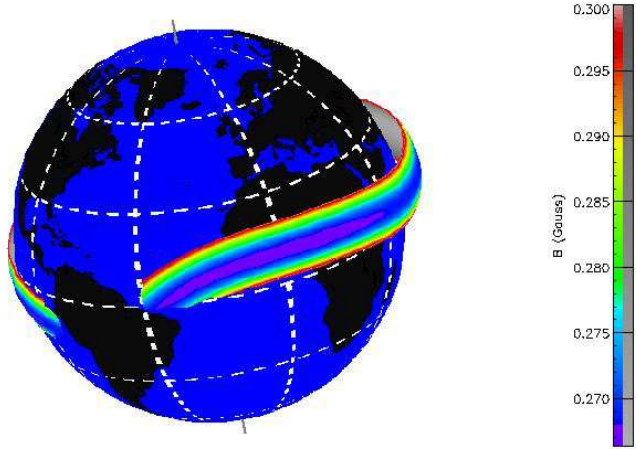,width=0.55\textwidth,clip=}
\caption{ Example of geometrical surface of a drift shell in a quasi-trapping belt. Notice the typical not-closed structure, in the vicinity of the SAA (SPENVIS package \cite{TREND}).}
\label{fig:belt}
\end{center}
\end{figure}

At higher energies the existing data come from measurements carried out by the Moscow 
Engineering Physics Institute. These data, taken at
altitudes ranging from 300$\div$1000 km with different instruments placed on
satellites and the Mir station [\cite{Voronov87}, \cite{Galper86}, \cite{Akimov87}],
established the existence of O(100 MeV) trapped leptons both in the Inner Van Allen Belts ({\it stably} trapped)
and in the region below ({\it quasi-trapped}), and  
determined their charge composition [\cite{Averin88}, \cite{Galper97}].
At these altitudes, the shell structure is strongly distorted in the vicinity of the SAA, and consequently
the observations are sensitive to different regions of trapped particles: the
Inner Van Allen belts over the SAA and quasi-trapping belts outside of the SAA. 
An example of the shell structure relevant at these altitudes is shown in Fig.\ref{fig:belt}: the shell evolves essentially above the atmosphere which it intercepts around the SAA.

The Russian measurements concern mainly the region of the SAA; very little data is available
at the corresponding altitudes outside the SAA. 
The measured ratio of $e^{+}$ to $e^{-}$ is found to depend strongly on the observed
population type. In the SAA, electrons dominate the positrons by a factor $\sim$ 10, a ratio similar
to that observed for the cosmic fluxes, while outside the SAA the two fluxes are similar and comparable to the 
$e^{+}$ flux inside the SAA [\cite{Galper97}]. 
However, the situation is not completely clear, since other groups report
a lower $e^-$ excess ($\sim$ 2) for the SAA [\cite{Kurnosova91}].

In the following, we use the high statistics data sample collected by the AMS experiment in 1998, 
for a detailed study
of the under-cutoff lepton fluxes in the O(1 GeV) energy range. The data are analyzed
in terms of the ca\-no\-ni\-cal invariant coordinates characterizing
the particle motion in the magnetic field: the L shell parameter, 
the equivalent magnetic equatorial radius of the shell,  
the equatorial pitch angle, $\alpha_{0}$, of the momentum $\vec{p}$ with the $\vec{B}$ field, 
and the mirror field $B_{m}$ at which the motion reflection occurs during bouncing [\cite{McIlwain61}, \cite{Hilton71}].

\section{AMS and the STS-91 flight}

The Alpha Magnetic Spectrometer (AMS), equip\-ped with a double-sided silicon microstrip tracker, 
has an analyzing power of $BD^{2}$=0.14 $Tm^{2}$, where {\it B} is the magnetic field intensity and 
{\it D} the typical path length in the field. 
A plastic scintillator time-of-flight system measures the particle velocity 
and an Aerogel Threshold Cerenkov counter provides the 
discrimination between proton and $e^{\pm}$. 
In the present analysis, a fiducial cone with a $28^{\mathrm{o}}$ half-angle opening aperture 
was defined to select the 
leptons entering the detector, re\-sul\-ting in an average acceptance of $\sim$160 $cm^{2}$sr.
Further details on the detector performance, lepton selection and background estimation can be found 
in \cite{Alcaraz00} and references therein.

The AMS was operated on the shuttle Discovery during a 10-day flight, beginning on June 2, 1998
(NASA mission STS-91). 
The detector, which was not magnetically
stabilized, recorded data during 17, 6, 7, and 14 hours 
pointing respectively
at 0$^o$, 20$^o$, 45$^o$, and 180$^o$ from the local zenith direction.
The results presented here are obtained from the data
of these periods. 
The orbital inclination was 51.7$^o$ in geographic coordinates, at a 
geodesic altitude of 370$\div$390 Km. Trigger rates varied between 100 and
700 Hz. The data from the SAA is excluded in our analysis.

The shuttle position and the AMS orientation in geographic
coordinates were provided continuously during the flight by the telemetry data. 
The values of L, \(\alpha_{0}\) and \(B_{m}\) of the detected 
leptons were calculated using the UNILIB package [\cite{TREND}] with a  
realistic magnetic field model, including both the internal  and the external contributions [\cite{IGRF}, 
\cite{Tsyganenko82}].

The AMS Field of View (FoV) in the (L,\(\alpha_{0}\)) coordinate space is determined both by the orbit parameters (geographic locations and flying attitude) 
and the finite acceptance of the detector. 

\begin{figure}[tbh]
\epsfig{file=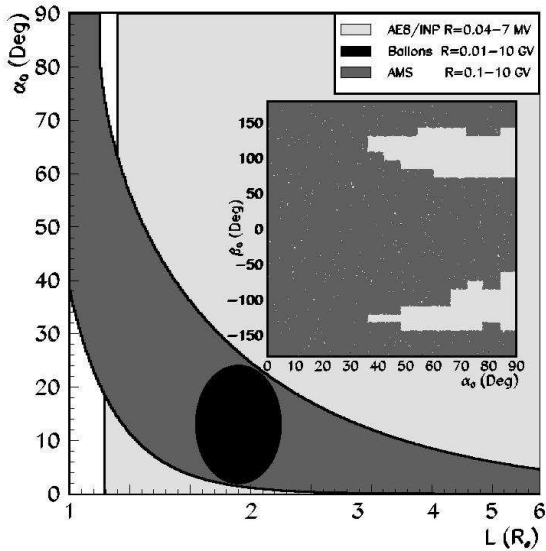,width=0.5\textwidth,clip=}
\caption{ Comparison of the field of view of AMS with balloons and satellite measurements in (L,$\alpha_{o}$). 
In the insert plot, the AMS coverage in $\beta_{o}$ vs $\alpha_{o}$ is shown.}
\label{fig:fov}
\end{figure}

A simulation was developped to determine the AMS FoV along the orbit and evaluate the effects due to the finite detector acceptance.
The results are shown in Fig.\ref{fig:fov}.


The \( (\alpha _{0},L) \) coverage is similar for the different attitudes. The finite acceptance of the detector influences 
essentially the definition of the lower contour.
The upper limit is imposed by the orbital altitude  and is 
described by the relation \( \sin \alpha _{0}=\sqrt{0.311/L^{3}B_{m}} \),
where \( B_{m}=0.225G \) is the minimum mirror field encountered along the AMS
orbit.     
Since the particles which are mirroring above the AMS altitude cannot
be observed, particles with large equatorial pitch angles can only
be detected at low L values (L$\leq$1.2). At larger L, only particles
with a smaller \( \alpha _{0} \) can be observed.
Because of the fixed flight attitudes, the azimuthal $\beta_{0}$ coverage in the local magnetic reference frame 
($\hat{z}$=$\hat{B}$, $\hat{x}$=$(\widehat{ \vec\nabla{B}})_{\perp}$, $\hat{y}$=$\hat{z}\times\hat{x}$) was not complete, as shown in the insert plot of Fig.\ref{fig:fov}.

\section{Data Analysis }

\begin{figure}[tbh]
\epsfig{file=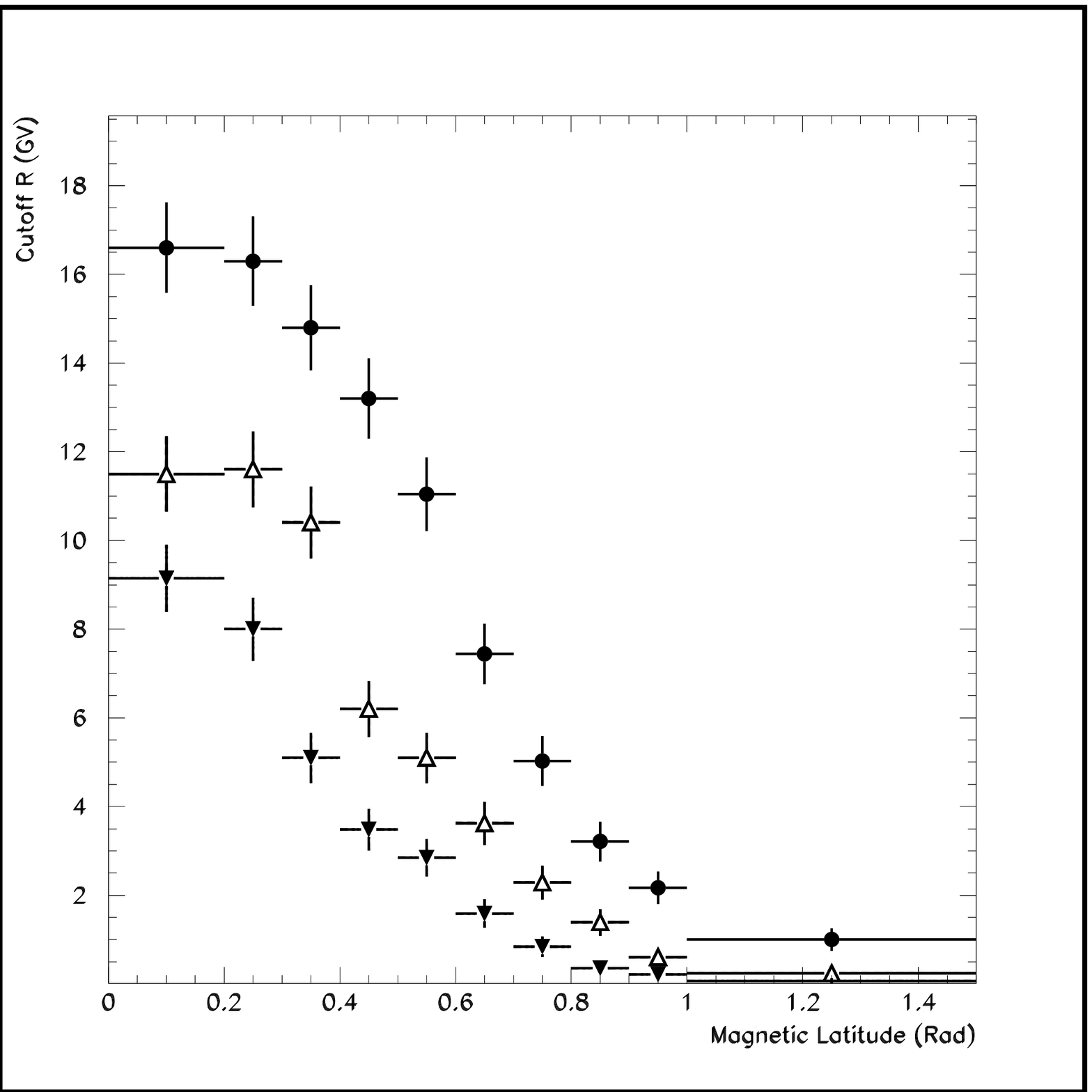,width=0.55\textwidth,clip=}
\caption{ 
 Effective cutoff as function of the magnetic latitude (filled triangles), defined as the lowest rigidity 
below which no primaries are found.
The highest rigidities above which all the leptons are primaries (points) and $50\%$ are secondaries (empty triangles) are also shown.
}
\label{fig:effcut}
\end{figure}

To reject the cosmic component of the measured lepton fluxes, the lepton trajectories
in the Earth's magnetic field were traced using a 4$^{th}$ order Runge Kutta method with adap\-tive step size. 
The equation of the motion was solved numerically and a particle 
was classified as trapped if its trajectory reached an altitude of 40 km [\cite{Alcaraz00}], 
taken as the dense atmosphere limit where the total probability of interaction is 50$\%$,
before its detection in AMS. 
Although satisfactory in most cases, this approach is less stable when the 
particle rigidity falls in the penumbra region, close to the cutoff value.
In this case, the trajectories become chaotic and small uncertainties 
in the reconstructed rigidity and in the B field can lead to a misclassification.
The validity of the adiabatic approach requires the 
parameter $\varepsilon =\rho /R $ to be small 
[\cite{Il'in86}, \cite{Schulz97}], where $\rho$ is
 the equatorial Larmor radius of a particle and $R$ the field radius of curvature at the equator. 
\cite{Il'in86}] shows that
the motion becomes cahotic if $\varepsilon \geq$ 0.1. The AMS data are consistent with this
limit even though the detected particle energies are relatively high.
To avoid such effects, we have defined an effective cutoff, $R_{eff}$, as the maximum 
rigidity value at a given magnetic latitude $\theta_{m}$ for which no traced lepton was found to be 
of cosmic origin. The $R_{eff}$ values as function of the magnetic latitude are shown as filled triangles in Fig.~\ref{fig:effcut}. 
We rejected from our sample all particles with $R\!>\!R_{e\!f\!f}$.

The residence time, $T_{f}$, of the under-cutoff particles is computed,
i.e. the total time spent by each particle in its motion above the atmosphere, before and after detection. 
The geographical location where the trajectories
intercept the atmosphere determine the lepton's {\it production} and {\it impact} points, 
defined as the position from which the particle leaves or enters the atmosphere.

\begin{figure}
\epsfig{file=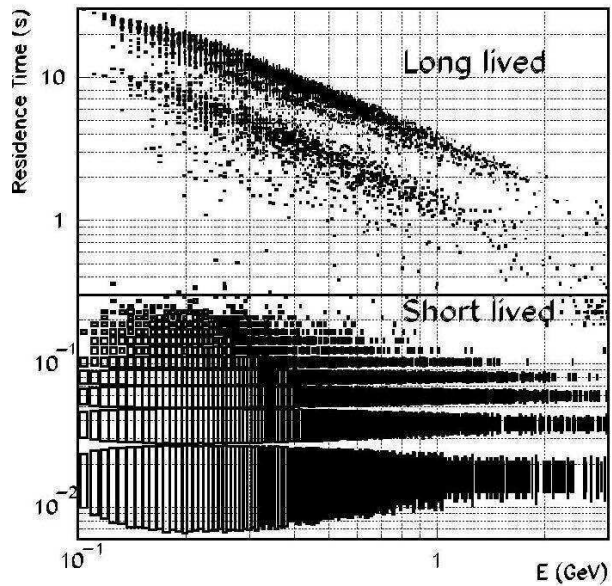,width=0.55\textwidth,clip=}
\caption{ Residence time vs energy for $e^{+}$. The same structure is observed for $e^{-}$.}
\label{fig:times}
\end{figure}
\begin{figure}
\epsfig{file=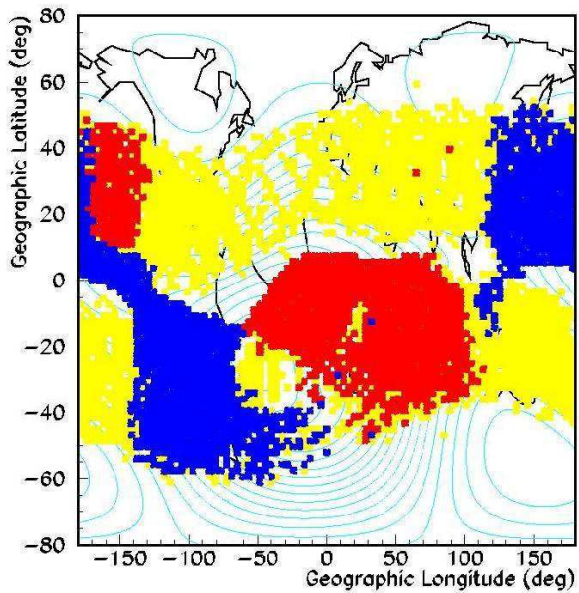,width=0.55\textwidth,clip=}
\caption{Geographical positions of production and impact points in the 
atmosphere. Yellow bands show the distribution for short-lived $e^{-}$,
red/blue bands show the production/impact distribution for long-lived $e^{-}$. 
A similar but complementary structure is observed for \( e^{+} \).}
\label{fig:traced}
\end{figure}

The residence time distribution as a function of 
energy is shown for positrons in Fig.\ref{fig:times}; the same behaviour
is observed for electrons. All observed leptons 
have residence times below $\sim$ 30 s, with 52$\%$ of the $e^{-}$ 
and 38$\%$ of the $e^{+}$ having a $T_{f}\!<\!0.3$ s 
independent of their e\-ner\-gy. The corresponding impact/production points are spread, for
both $e^{+}$ and $e^{-}$, over two bands on either side of magnetic equator, as indicated by 
the the yellow bands in Fig.\ref{fig:traced}.

A scaling law, $T_{f} \approx E^{-1}$, is observed for the remaining 
leptons: they are disposed in two diagonal bands separated by a 
difference in $T_{f}$ of $\approx 2.2/E$ s. The impact/production points for $e^{+}$ 
are localized in the red/blue spots of Fig.\ref{fig:traced}: the same regions 
describe respectively the production/impact regions of $e^{-}$.

The data have been previously published by the AMS col\-la\-bo\-ra\-tion 
[\cite{Alcaraz00}] using the 
terminology of {\it short-lived} and {\it long-lived} to 
classify the particles with $T_{f}$ below and 
above 0.2 s. However, no interpretation was advanced at that time to describe 
the observed distributions.
\cite{Lipari01}] has discussed the AMS results qualitatively. 

An exhaustive explanation must take into account the geometry of the shells 
relevant to the AMS measurements and the fact that these 
shells evolve partially under the atmosphere; therefore, no permanent trapping 
can occur. The residence times are determined by the 
periodicities of the drift ($\tau_{d}$) and boun\-cing ($\tau_{b}\ll \tau_{d}$)  motions, the type of motion which dominates
depends on the relative fraction of the shell mirror points lying above the atmosphere.

\begin{figure}
\epsfig{file=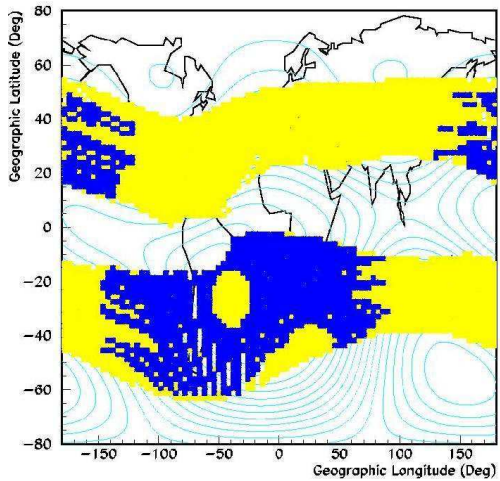,width=0.45\textwidth,clip=}
\label{fig:shellplot}
\vskip -0.45truecm
\caption{ Distribution of intersection points with atmosphere for the
drift shells crossed by AMS. Yellow region corresponds to shells with 
$\mathrm{B_{m}}\geq 0.48\, \mathrm{L}^{0.41}$ G, blue region to 
$\mathrm{B_{m}}\leq 0.48\, \mathrm{L}^{0.41}$ G.  }
\end{figure}

The impact/production points correspond to the intersection of the shell surfaces with the atmosphere, 
as shown in Fig.\ref{fig:shellplot}, where particles generated in interactions are injected into the shells.
Long-lived and short-lived particles move along shells with different values of \( B_{m} \) or,
equivalently $\alpha_{0}$, which determine the mirror height on each field line. 
For high $B_{m}$ values, or low $\alpha_{0}$, the mirror height is very low and the shells penetrate
into the atmosphere at nearly all longitudes. Therefore particles are absorbed shortly after injection in the shells. 
This is shown by the yellow bands in Fig.\ref{fig:shellplot} 
corresponding to shells with $\mathrm{B_{m}}\geq 0.48\, \mathrm{L}^{0.41}$ Gauss; they reproduce well 
the impact/production points for short-lived leptons.
When $B_{m}$ is lower, or $\alpha_{0}$ closer to $90^{\mathrm{o}}$, the shells descend below the 
atmosphere only in the vicinity of the SAA, indicated by the blue regions in Fig.\ref{fig:shellplot}, which correspond
to shells with $\mathrm{B_{m}}\leq 0.48\, \mathrm{L}^{0.41}$ Gauss, and reproduce the impact/production points of the long-lived leptons.
These particles can drift nearly an entire revolution before absorption in the atmosphere.
For the short-lived component, the bouncing motion is dominant; 
the residence time is given by $T_{f}=k\tau_{b}$, where $\tau_{b}$ is 
the bouncing motion period, and {\it k} is an integer or half-integer between $1/2\leq~k~\leq~5$. 
In the dipolar field model, $\tau_{b}$ is given 
by $\tau_{b}=f(\alpha_{o})L$,
where {\it f} is a slowly varying function of $\alpha_{o}$; 
the upper limit is $\tau\sim 300~ms$ for the AMS data. The sub-structure seen in the short-lived component of Fig.\ref{fig:times} 
is due to the discrete values of k and the different L shells crossed during the AMS orbits. 
For the long-lived component,
the drift motion is dominant and $T_{f}=k'~\tau_{d}$, where $\tau_{d}$ is the drift motion period, 
$\tau_{d}=f'(\alpha_{o})/EL$, where {\it f'} is a slowly varying function of $\alpha_{o}$, and {\it k}' is
a number less than one, corresponding to the fraction of a complete drift shell spanned 
by a particle. The two bands seen for the long-lived component of Fig.\ref{fig:times}
correspond to fractions of $\sim$0.65 and $\sim$0.25 of a complete drift.

\section{AMS Results}

For the description of under-cutoff fluxes, the energy E, the L parameter and the equatorial pitch angle 
$\alpha_{0}$ were used (this is preferred to $B_{m}$ since 
limited to $0^{\mathrm{o}}\div 90^{\mathrm{o}}$).
A three-dimensional grid (E, L, $\alpha_{0}$) was defined to build flux maps. 
A linear binning in $\alpha_{0}$ and
logarithmic variable size for L and E bins were choosen to optimize the statistics in each bin.
The interval limits and bin widths are listed in Table 1.

\begin{figure} 
\epsfig{file=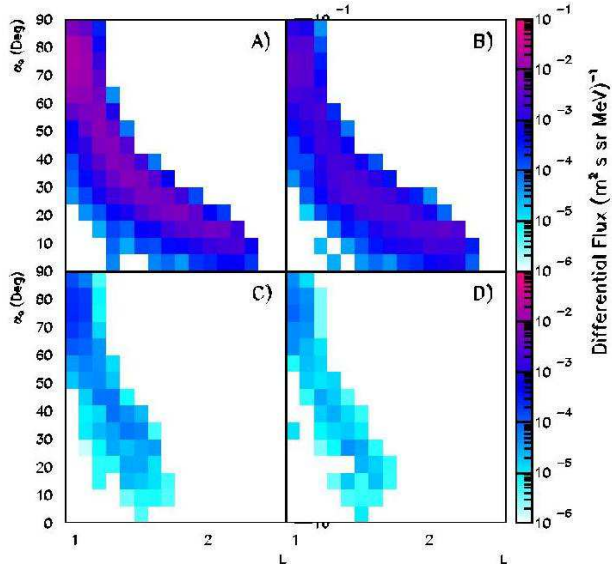,width=0.55\textwidth,clip=}
\caption{Flux maps for two different energy bins: A), B) $e^{+}$, $e^{-}$ between 0.315$\leq$E$\leq$0.486 GeV 
and C), D) $e^{+}$, $e^{-}$ between 1.77$\leq$E$\leq$2.73 GeV. }
\label{fig:map}
\end{figure}

The flux maps in (L, $\alpha_{0}$) at constant E give the distribution of particle populations 
at the altitude of AMS.
Nine maps at constant E have been made. Two different maps for two different 
energy bins of $e^{+}$ and $e^{-}$ are shown in Fig.\ref{fig:map}.

The flux is limited by the cutoff rigidity $R_{c}$: on a given shell only particles with
R $\leq R_{c}$ are allowed to populate the shell, hence
lower and lower energy particles populate higher and higher shells.

\begin{figure}
\epsfig{file=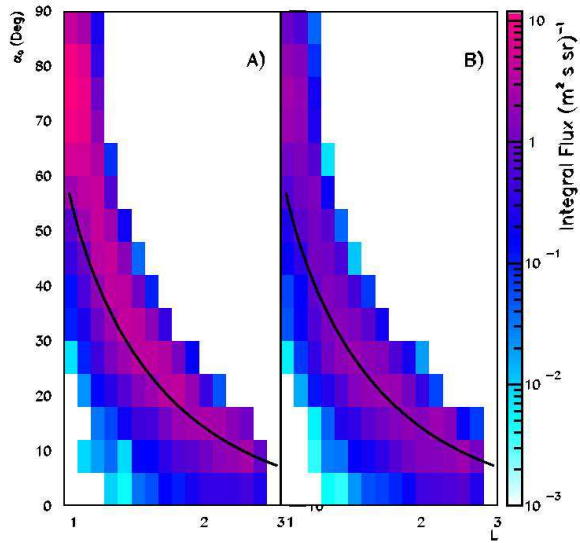,width=0.485\textwidth,clip=}
\caption{ The integrated flux maps for $e^{+}$ (A) and $e^{-}$ (B) between 0.205$\leq$E$\leq$2.73 GeV. 
The line shows the curve below which no {\it quasi-trapped} leptons are found.}
\label{fig:intmap}
\end{figure}

\begin{figure} 
\epsfig{file=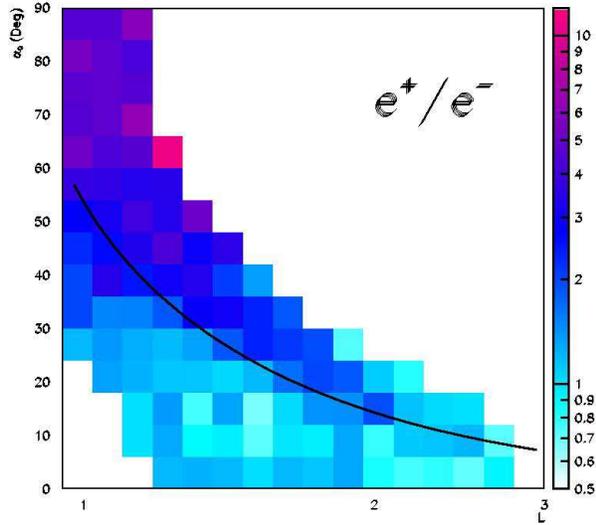,width=0.5\textwidth,clip=}
\caption{The integrated $e^{+}$/$e^{-}$ ratio  between 0.205$\leq$E$\leq$2.73 GeV.
The line shows the curve below which no {\it quasi-trapped} leptons are found.}
\label{fig:qratio}
\end{figure}

The $e^{+},\,e^{-}$ flux maps and their ratio in the energy interval 0.2$\div$2.7 GeV are shown 
in Fig. \ref{fig:intmap} and Fig. \ref{fig:qratio} respectively. 
The solid line in the two plots identifies the lower boundary in (L,$\alpha_{o}$) below which no leptons
can be found with residence times larger than 0.3 s and is defined by the relation $\sin\alpha_{c}=0.8L^{-1.7}$. 

\begin{figure} 
\epsfig{file=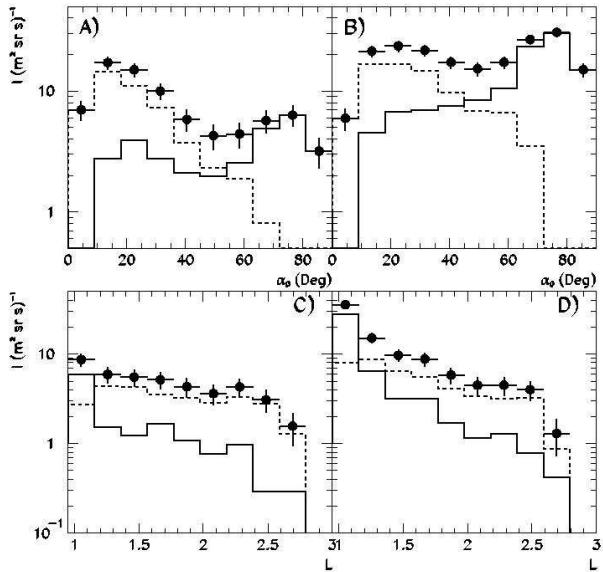,width=0.5\textwidth,clip=}
\caption{The integrated flux as function of $\alpha_{0}$ and as function of L for $e^{-}$ (A, C) and $e^{+}$ 
(B, D)  between 0.205$\leq$E$\leq$2.73 GeV. The full line shows the long-lived component,
the dashed line shows the short-lived component, while the points show the total flux. }
\label{fig:int1d}
\end{figure}

Above the curve, for increasing values
of $\alpha_{o}$, the long-lived component of the fluxes becomes increasingly dominant. This is demonstrated 
in Fig. \ref{fig:int1d} where the same distributions, integrated over $\alpha_{o}$ (C,D) and L (A,B), 
are shown. The contributions of leptons with $T_{f}\!<\!0.3$ s and $T_{f}\!>\!0.3$ s are represented 
with dashed and 
solid lines respectively. Above $\alpha_{o}\!>\!60^{\mathrm{o}}$ the flux is due substantially to the 
long-lived component; 
the $e^{+}$ flux represents $\approx 80\%$ of the total leptonic flux, while at the
same level or less than the $e^{-}$ flux in the low $\alpha_{o}$ region. The long-lived component 
dominates only at very low L values where the positron excess is more pronounced.

\begin{figure}
\epsfig{file=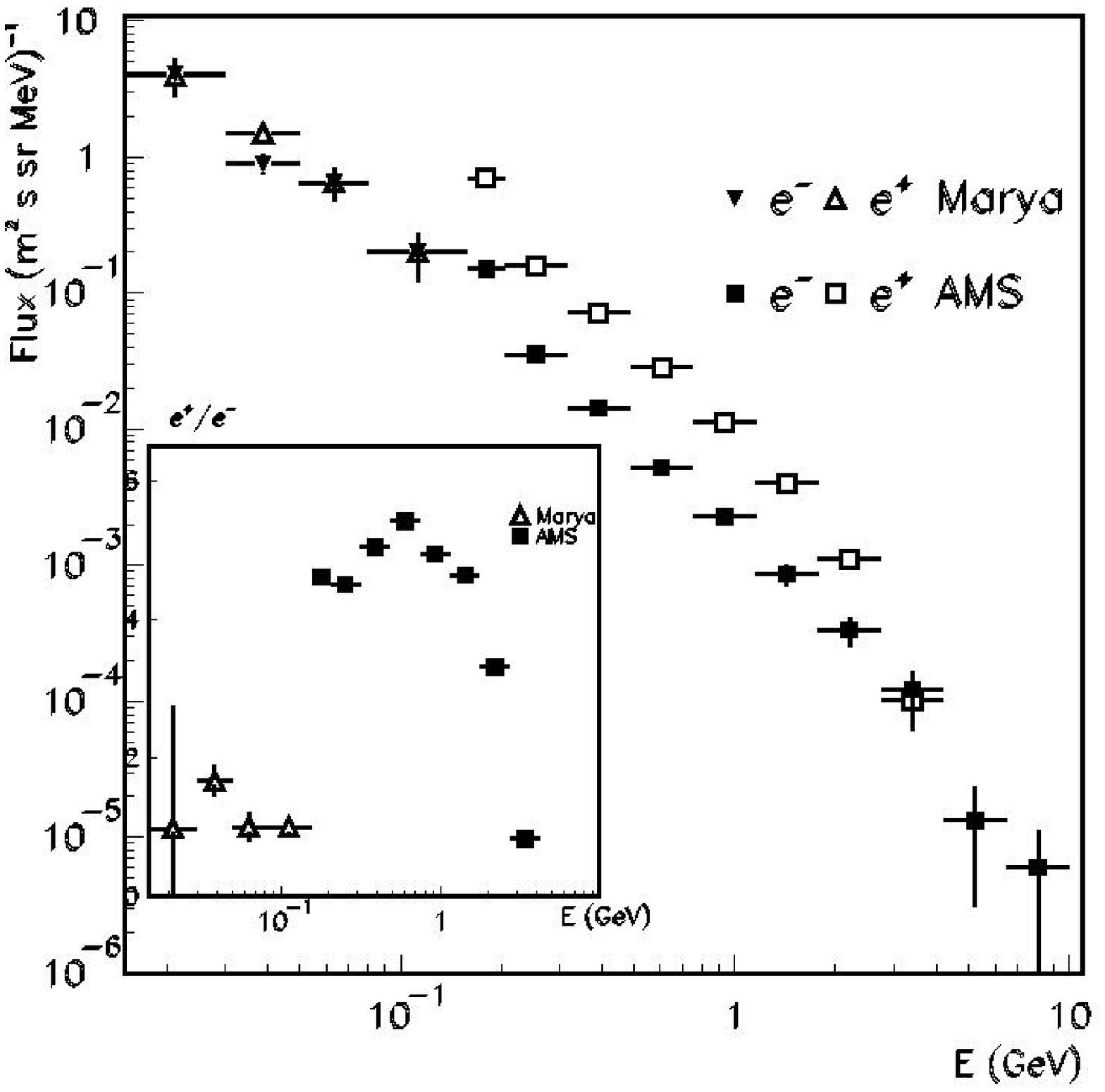,width=0.45\textwidth}
\caption{Energy spectrum comparison between AMS and Marya for $e^{+}$ and $e^{-}$ for particles 
with $\alpha_{0} \geq 70^{\mathrm{o}}$. In the insert plot, the  $e^{+}$/ $e^{-}$ ratio comparison is shown.}  
\label{fig:marya}
\end{figure}

This is seen clearly in the energy spectra for particles with $\alpha_{0}\!\geq\! 70^{\mathrm{o}}$, 
shown in Fig. \ref{fig:marya}, which is superimposed with the lower energy
 measurements from MARYA [\cite{Galper97}]. 
At large pitch angles, the $e^{+}$ flux is higher than $e^{-}$ flux by a factor $\sim$ 4.5, 
in contrast with MARYA data which indicate the same level of flux for both lepton charges.
The critical pitch angle $\alpha_{c}$ can explain the presence of the two well separated components 
in the residence time: the albedo (short-lived) and quasi-trapped (long-lived) ones.
Particles inside a cone with a half-opening angle $\alpha_{c}$ around $\vec{B}$ will enter the atmosphere every bounce and therefore will disappear rapidly, 
while those outside it will enter the atmosphere only near SAA. In this context, $\alpha_{c}$ can be
defined as the {\it equatorial bouncing loss cone angle}, i.e. the largest pitch angle for which particles enter the atmosphere every bounce.
Taking into account $\alpha_{c}$, the residence time
 can be written as $T_{f}=k\tau_{b}\theta(-\alpha_{o}+\alpha_{c})+k'\tau_{d}\theta(\alpha_{o}-\alpha_{c})$, where $\theta$ is the
Heavyside step function. 
The two terms correspond to very different components of motion ($\tau_{b}\ll\tau_{d}$), according
to the bounce loss cone on that field line.
Furthermore, all the observed particles are in a drift loss cone angle since all of them enter the atmosphere within one revolution after injection. This explains
the absence of a peak at $90^{o}$ in the pitch angle distributions of  Figs. \ref{fig:int1d} A and B.

\section{Discussion}

The AMS data establish 
the existence of leptonic radiation belts, with particle energies in the range of se\-ve\-ral GeV,
below the Inner Van Allen belts.
The particles populating these belts are not stably trapped since the corresponding
drift shells are not closed over above the atmosphere in the region of the SAA. 

At any given L, a critical value of the equatorial pitch angle, $\alpha_{c}$, the {\it bouncing loss cone}, can be defined to 
distinguish the long-lived, or {\it quasi-trapped}, and the short-lived, or {\it albedo}, components of the fluxes.
The same value is found to se\-pa\-ra\-te the regions where the $e^{+}/e^{-}$ ratio is above or around 
unity: the charge composition shows a clear dominance of positively charged leptons in a
 definite region of the (L,$\alpha_{o}$) space above $\alpha_{c}(L)$. 

The observed behaviour distinguishes these belts from the Inner Van Allen belts and limits 
the possible injection/loss mechanisms to those acting on a time scale much shorter than the 
typical particle re\-si\-dence times. Mechanisms related to Coulomb scat\-te\-ring,
like pitch angle diffusion, are ruled out since they imply much longer time scales.
Moreover, the observed charge ratio 
distribution provides an important constraint for potential models.

The interaction of primary cosmic rays and inner radiation belt protons with atmospheric nuclei
in the regions of shell intersection with the atmosphere are a natural mechanism for the production 
of secondary leptons through the $\pi-\mu-e$ or $\pi-\gamma-e$ decay chains. This
leads to a $e^{+}$ excess over $e^{-}$ and seems suitable to explain the observed charge 
ratio for the {\it quasi-trapped} flux [\cite{Voronov95a},\cite{Gusev82}]. However, for the {\it albedo} flux the charge ratio 
is of the order of unity, as seen in Fig. \ref{fig:qratio}, and other mechanisms might be present.

Recent Monte Carlo studies based on this mechanism have been able to fully reproduce the under-cutoff proton 
spectrum reported by AMS [\cite{Derome00}], while a less good agreement for the under-cutoff 
lepton spectrum [\cite{Derome01}] was obtained. In \cite{Lipari01}] the influence of
geomagnetic effects, mainly related to the East-West asymmetry for cosmic protons, is taken into 
account to qualitatively explain the observed charge ratio. However, more refined studies are needed to
definitely exclude contributions from other mechanisms, i.e. acceleration processes acting on the leptons resulting from
the decays of $\beta$-active secondary nuclei and neutrons of albedo and solar 
origin [\cite{Voronov95b}]. 

In conclusion, the AMS under-cutoff lepton spectrum 
can be described naturally in terms of the canonical adiabatic variables as\-so\-cia\-ted with the E\-arth's
magnetic field taking into account the role played by the atmosphere. There are clear indications 
that $\pi$ decays can account for the {\it quasi-trapped} component of the flux, while the situation is less clear for the
{\it albedo} component where other processes may contribute.

\section{Acknowledgements}   

We gratefully acknowledge our collegues in AMS, 
in particular Z. Ren and V. Choutko. We are also grateful to P.Lipari for useful discussions on the interpretation of the AMS data. We greatly benefit of
the software libraries (UNILIB,SPENVIS) developed in the context of the Trapped Radiation ENvironment Development (TREND) project for ESTEC, and we
thank D.Heynderickx for his help.
\vskip 0.5cm
{\it This work has been partially supported by Italian Space Agency (ASI) under the contract ARS 98/47.}

\end{document}